\documentstyle[12pt,epsf]{article}
 \newcommand{\insertplot}[5]{\begin{figure}
 \hfill\hbox to 0.05in{\vbox to #5in{\vfill
 \inputplot{#1}{#4}{#5}}\hfill}
 \hfill\vspace{-.1in}
 \caption{#2}\label{#3}
 \end{figure}}
 \newcommand{\inputplot}[3]{
 \special{ps: plotfile #1}
}

\begin{document}

\title{STATIC BLACK HOLE SOLUTIONS 
WITH AXIAL SYMMETRY}
\vspace{1.5truecm}
\author{
{\bf Burkhard Kleihaus, and Jutta Kunz}\\
Fachbereich Physik, Universit\"at Oldenburg, Postfach 2503\\
D-26111 Oldenburg, Germany}

\vspace{1.5truecm}


\maketitle
\vspace{1.0truecm}

\begin{abstract}
We construct a new class of asymptotically flat black hole solutions
in Einstein-Yang-Mills and Einstein-Yang-Mills-dilaton theory.
These black hole solutions are static, 
and they have a regular event horizon.
However, they possess only axial symmetry.
Like their regular counterparts,
the black hole solutions are characterized by two integers,
the winding number $n$
and the node number $k$ of the gauge field functions.                          
\end{abstract}
\vfill
\noindent {Preprint hep-th/9704060} \hfill\break
\vfill\eject

\section{Introduction}

The ``no hair'' conjecture for black holes states,
that black holes are completely characterized 
by their mass $M$, their charge $Q$ and their angular momentum $J$.
This conjecture presents a generalization of
rigorous results obtained for scalar fields coupled to gravity
\cite{bek} as well as 
for Einstein-Maxwell (EM) theory \cite{hair-em}.
In EM theory,
the unique family of stationary Kerr-Newman black holes
with nontrivial values of $M$, $Q$, and $J$
contains the stationary Kerr black holes for $Q=0$,
the static Reissner-Nordstr\o m black holes for $J=0$
and the static Schwarzschild black holes for $J=Q=0$.
Notably, the static black hole solutions in EM theory
are spherically symmetric, and the stationary black holes
are axially symmetric.

In recent years counterexamples to the ``no hair'' conjecture
were established in various theories with non-abelian fields,
including Einstein-Yang-Mills (EYM) theory,
Einstein-Yang-Mills-dilaton (EYMD) theory,
Einstein-Yang-Mills-Higgs (EYMH) theory,
and Einstein-Skyrme (ES) theory.
Apart from some perturbative solutions \cite{ewein,slomo},
these non-abelian black holes are all static and
spherically symmetric. They are asymptotically flat
and possess a regular event horizon.
And in particular these black hole solutions
possess non-trivial matter fields outside
the event horizon.

In SU(2) EYM theory, for instance,
there exists in addition to the Schwarzschild solution
a whole sequence of neutral static spherically symmetric
non-abelian black hole solutions,
for arbitrary radius $X_{\rm H}$ of the event horizon \cite{eym}.
Within this sequence, the solutions are labelled by
the node number $k$ of the single gauge field function.
These black hole solutions have regular counterparts,
the Bartnik-McKinnon solutions \cite{bm},
obtained in the limit $X_{\rm H} \rightarrow 0$.
Both, black hole solutions and regular solutions
are unstable \cite{strau}.

The situation is analogous in EYMD theory \cite{eymd}.
Here the dilaton coupling constant $\gamma$ represents a parameter;
for $\gamma = 1$ contact with the low energy effective action
of string theory is made,
whereas in the limit $\gamma \rightarrow 0$ 
the dilaton decouples and EYM theory is obtained.
In contrast, EYMH and ES theory allow for stable static
spherically symmetric black hole solutions with hair.

Here we present strong (numerical) evidence,
that static black hole solutions in EYM and EYMD theory
need not be spherically symmetric. Indeed, the 
asymptotically flat black hole solutions
we construct possess only axial symmetry.
They are characterized by two integers,
the winding number $n>1$
and the node number $k$ of the gauge field functions.                          
Constructing their regular counterparts,
we have conjectured their existence previously \cite{kk1,kk2}.

\section{\bf Static axially symmetric ansatz}

We consider the SU(2) Einstein-Yang-Mills-dilaton action
\begin{equation}
S=\int \left ( \frac{R}{16\pi G} + L_M \right ) \sqrt{-g} d^4x
\   \end{equation}
with
\begin{equation}
L_M=-\frac{1}{2}\partial_\mu \Phi \partial^\mu \Phi
 -e^{2 \kappa \Phi }\frac{1}{2} {\rm Tr} (F_{\mu\nu} F^{\mu\nu})
\ , \end{equation}
$F_{\mu \nu} = 
\partial_\mu A_\nu -\partial_\nu A_\mu + i e \left[A_\mu , A_\nu \right] $,
and $e$ and $\kappa$ are the Yang-Mills and dilaton coupling constants,
respectively.

The static axially symmetric ans\"atze for the metric
and the matter fields of the black hole solutions
agree with those of the corresponding regular solutions \cite{kk2}.
Here we parametrize the ans\"atze from the beginning
in terms of the coordinates $r$ and $\theta$ 
(instead of $\rho=r \sin \theta$, $z=r \cos \theta$ \cite{kk2}).
The metric then reads in isotropic coordinates
\begin{equation}
ds^2=
  - f dt^2 +  \frac{m}{f} d r^2 + \frac{m r^2}{f} d \theta^2 
           +  \frac{l r^2 \sin^2 \theta}{f} d\phi^2
\ , \label{metric} \end{equation}
with $f$, $m$ and $l$ being only functions of $r$ and $\theta$.
We parametrize the purely magnetic gauge field ($A_0=0$) 
by \cite{rr,kk1,kk2}
\begin{equation}
A_\mu dx^\mu =
\frac{1}{2er} \left[ \tau^n_\phi 
 \left( H_1 dr + \left(1-H_2\right) r d\theta \right)
 -n \left( \tau^n_r H_3 + \tau^n_\theta \left(1-H_4\right) \right)
  r \sin \theta d\phi \right]
\ , \label{gf1} \end{equation}
with the Pauli matrices $\vec \tau = ( \tau_x, \tau_y, \tau_z) $ and
$\tau^n_r = \vec \tau \cdot 
(\sin \theta \cos n \phi, \sin \theta \sin n \phi, \cos \theta)$,
$\tau^n_\theta = \vec \tau \cdot 
(\cos \theta \cos n \phi, \cos \theta \sin n \phi, -\sin \theta)$,
$\tau^n_\phi = \vec \tau \cdot (-\sin n \phi, \cos n \phi,0)$.
We refer to $n$ as the winding number of the solutions.
Again, the four gauge field functions $H_i$ \cite{foot1}
and the dilaton function $\Phi$ depend only on $r$ and $\theta$.
For $n=1$ the spherically symmetric ansatz of ref.~\cite{eymd} 
is recovered with $H_1=H_3=0$, $H_2=H_4=w(r)$ and $\Phi=\Phi(r)$.

Denoting the stress-energy tensor of the matter fields by $T_{\mu}^{\nu}$,
with this ansatz the energy density $\epsilon =-T_0^0=-L_M$ becomes
\begin{eqnarray}
-T_0^0 = & \frac{f}{2m} \left[
 (\partial_r \Phi )^2 + \frac{1}{r^2} (\partial_\theta \Phi )^2 \right]
       + e^{2 \kappa \Phi} \frac{f^2}{2 e^2 r^4 m} \left\{
 \frac{1}{m} \left(r \partial_r H_2 + \partial_\theta H_1\right)^2 
 \right.
\nonumber \\
      & +  \left.
   \frac{n^2}{l} \left [
  \left(  r \partial_r H_3 - H_1 H_4 \right)^2
+ \left(r \partial_r H_4 + H_1 \left( H_3 + {\rm ctg} \theta \right)
    \right)^2 \right. \right.
\nonumber \\
      & +  \left. \left.
  \left(\partial_\theta H_3 - 1 + {\rm ctg} \theta H_3 + H_2 H_4
     \right)^2 +
  \left(\partial_\theta H_4 + {\rm ctg} \theta \left( H_4-H_2 \right) 
   - H_2 H_3 \right)^2 \right] \right\}
\ , \label{edens} \end{eqnarray}
where the gauge field terms in the first, second and third line
derive from $F_{r\theta}$, $F_{r\phi}$ and $F_{\theta\phi}$,
respectively.

The system possesses a residual abelian gauge invariance 
\cite{kkb,kk,kk1,kk2}.
With respect to the transformation
\begin{equation}
 U= e^{i\Gamma(r,\theta) \tau^n_\phi}
\  \label{gauge} \end{equation}
the functions $H_1$ and $H_2$ transform inhomogeneously
like a 2-dimensional gauge field 
($H_1 \rightarrow H_1 - 2 r \partial_r \Gamma$,
 $H_2 \rightarrow H_2 + 2  \partial_\theta \Gamma$),
whereas $(H_3+{\rm ctg} \theta, H_4)$ transforms like a scalar doublet.
To fix the gauge we choose the
same gauge condition as previously \cite{kkb,kk,kk1,kk2}
\begin{equation}
 r \partial_r H_1 - \partial_\theta H_2 = 0 
\ . \label{gc1} \end{equation}

With the ansatz (\ref{metric})-(\ref{gf1})
and the gauge condition (\ref{gc1}) 
we then obtain the set of EYMD field equations \cite{kknew}.

\section{\bf Horizon and boundary conditions}

We are now looking for static axially symmetric solutions
of the field equations, which are asymptotically flat,
have a finite mass, and possess a regular event horizon.
The presence of the regular event horizon
is the essential new feature of the
static axially symmetric black hole solutions
with respect to the corresponding regular solutions \cite{kk2}.

The event horizon of the static black hole solutions
is characterized by $g_{tt}=-f=0$.
(In isotropic coordinates $g_{rr}$ is finite at the horizon.)
We now impose that the horizon of the black hole solutions
resides at a surface of constant $r$, $r=r_{\rm H}$
\cite{rh1,rh2}.
Requiring the horizon to be regular,
we obtain the following set of boundary conditions 
for the functions at the horizon ($r=r_{\rm H}$)
\begin{eqnarray}
&f=m=l=0 \ , \ \ \  \partial_r \Phi=0 \ ,  \ \ \
 \partial_\theta H_1 + r \partial_r H_2 = 0 \ ,
\nonumber \\
& r \partial_r H_3-H_1 H_4=0,   \ \ \ 
 r \partial_r H_4+H_1( H_3 + {\rm ctg} \theta) =0
\ . \label{bc3} \end{eqnarray}
Here the gauge field conditions in the first and second line
imply $F_{r\theta}=0$ and $F_{r\phi}=0$,
respectively.
Thus the equations of motion yield only three boundary conditions
for the four gauge field functions $H_i$;
one gauge field boundary condition is left indeterminate.
However, for the black hole solutions
precisely one free boundary condition at the horizon is necessary
to completely fix the gauge.
The reason is, that in contrast to the case 
of the regular solutions \cite{kk2}, the gauge condition (\ref{gc1})
still allows for non-trivial gauge transformations
for the black hole solutions,
which satisfy $r^2 \partial^2_r \Gamma
+r \partial_r \Gamma
+  \partial^2_\theta \Gamma = 0$.
We have implemented various choices of gauge
(e.g.~$\partial_r H_1 = 0$),
obtaining the same results for the gauge invariant quantities.

The boundary conditions at infinity and along the $\rho$- and $z$-axis
are the same as for the regular solutions \cite{kk2}
(when expressed in terms of the functions $F_i$ \cite{foot1}).
In terms of the functions $H_i$,
the boundary condition at infinity ($r=\infty$) are
\begin{equation}
f=m=l=1 \ , \ \ \ \Phi=0 \ , \ \ \
H_2=H_4=\pm 1, \ \ \ H_1=H_3=0
\ . \label{bc2} \end{equation}
For the gauge field functions they imply,
that the solutions are magnetically neutral.
(Any finite value of the dilaton field
can always be transformed to zero via
$\Phi \rightarrow \Phi - \Phi(\infty)$, 
$r \rightarrow r e^{-\kappa \Phi(\infty)} $.)
Along the $\rho$- and $z$-axis $H_1=H_3=0$.
The derivatives of all other functions 
with respect to $\theta$ vanish on these axes.

\section{\bf Mass and temperature}

Let us change to dimensionless quantities now;
the dimensionless coordinate $x=(e/\sqrt{4\pi G}) r$, 
the dimensionless dilaton function $\varphi = \sqrt{4\pi G} \Phi$ and  
the dimensionless dilaton coupling constant
$\gamma =\kappa/\sqrt{4\pi G}$
($\gamma=1$ corresponds to string theory).

The mass $M$ of the black hole solutions can be obtained directly from
the total energy-momentum ``tensor'' $\tau^{\mu\nu}$
of matter and gravitation,
$M=\int \tau^{00} d^3r$ \cite{wein}.
The dimensionless mass $\mu =(e/\sqrt{4\pi G}) G M$
is then determined by the derivative of the metric function $f$
at infinity,
\begin{equation}
\mu = \frac{1}{2} x^2 \partial_x f |_\infty
\ , \label{mass} \end{equation}
as for the regular solutions \cite{kk2}.
Similarly, the derivative of the dilaton function at infinity
determines the dilaton charge $D  =  x^2 \partial_x \varphi |_\infty $.

In order to define a temperature $T=\kappa_{\rm sg} /(2 \pi)$,
the surface gravity $\kappa_{\rm sg}$ \cite{ewein},
\begin{equation}
\kappa^2_{\rm sg}=-(1/4)g^{tt}g^{ij}(\partial_i g_{tt})(\partial_j g_{tt})
\ , \label{sg} \end{equation}
must be constant at the horizon of the black hole solutions.
Let us therefore consider the metric functions at the horizon.
Expanding the equations in the vicinity of the horizon
in powers of $x-x_{\rm H}$,
we observe, that the metric functions are quadratic
in $x-x_{\rm H}$,
$f(x,\theta)=f_2(\theta)(x-x_{\rm H})^2 + O(x-x_{\rm H})^3$, 
$m(x,\theta)=m_2(\theta)(x-x_{\rm H})^2 + O(x-x_{\rm H})^3$
and likewise for $l(x,\theta)$.
(Axial symmetry requires $m(x,\theta=0)=l(x,\theta=0)$.)
Consequently, we obtain the temperature
\begin{equation}
T=\frac{f_2(\theta)}{2 \pi \sqrt{m_2(\theta)} } 
\ , \label{temp} \end{equation}
which indeed is constant \cite{kknew}.

\section{\bf Numerical results}

Subject to the above boundary conditions,
we solve the equations for the black hole solutions numerically.
We map spatial infinity to the finite value $\bar{x}=1$
and employ the radial coordinate $\bar{x} = 1-(x_{\rm H}/x)$.
The equations are then discretized on a non-equidistant
grid in $\bar{x}$ and $\theta$.
Typical grids used have sizes $150 \times 30$, 
and cover the integration region 
$0\leq\bar{x}\leq 1$, $0\leq\theta\leq\pi/2$.
We employ the same numerical algorithm \cite{schoen}
as for the regular solutions \cite{kk2,kknew}.
The numerical error for the functions is estimated to be 
on the order of $10^{-3}$.

In Table~1 we show the dimensionless mass, temperature
and dilaton charge of the axially symmetric black hole
solutions of EYM theory ($\gamma=0$)
and EYMD theory (for $\gamma=1$)
with winding numbers $n\le3$, node numbers $k\le3$
and $x_{\rm H}=1$.

For all solutions the energy density 
$\epsilon$ (\ref{edens}) of the matter fields 
is angle-dependent at the horizon.
Considering small values of $x_{\rm H}$
the energy density of the black hole solutions
tends toward the energy density of the corresponding regular solutions.
However, the limit $x_{\rm H} \rightarrow 0$ of the energy density
is not smooth, because the angular dependence at the horizon 
remains.
(The limit is neither smooth for the spherical solutions with $n=1$.)
The reason is, that the magnetic field of the black hole solutions
is purely radial at the horizon, $\vec B= B_r \vec e_r$
($B_r=F_{\theta\phi}$),
because the boundary conditions (\ref{bc3}) require
$B_\theta=-F_{r\phi}=0=B_\phi=F_{r\theta}$.
In contrast, for the regular solutions the magnetic field
also has non-vanishing $B_\theta$ at the origin,
and the contributions from both $B_r$ and $B_\theta$
precisely add to an angle-independent density at the origin.

\begin{figure}
\centering
\epsfysize=11cm
\mbox{\epsffile{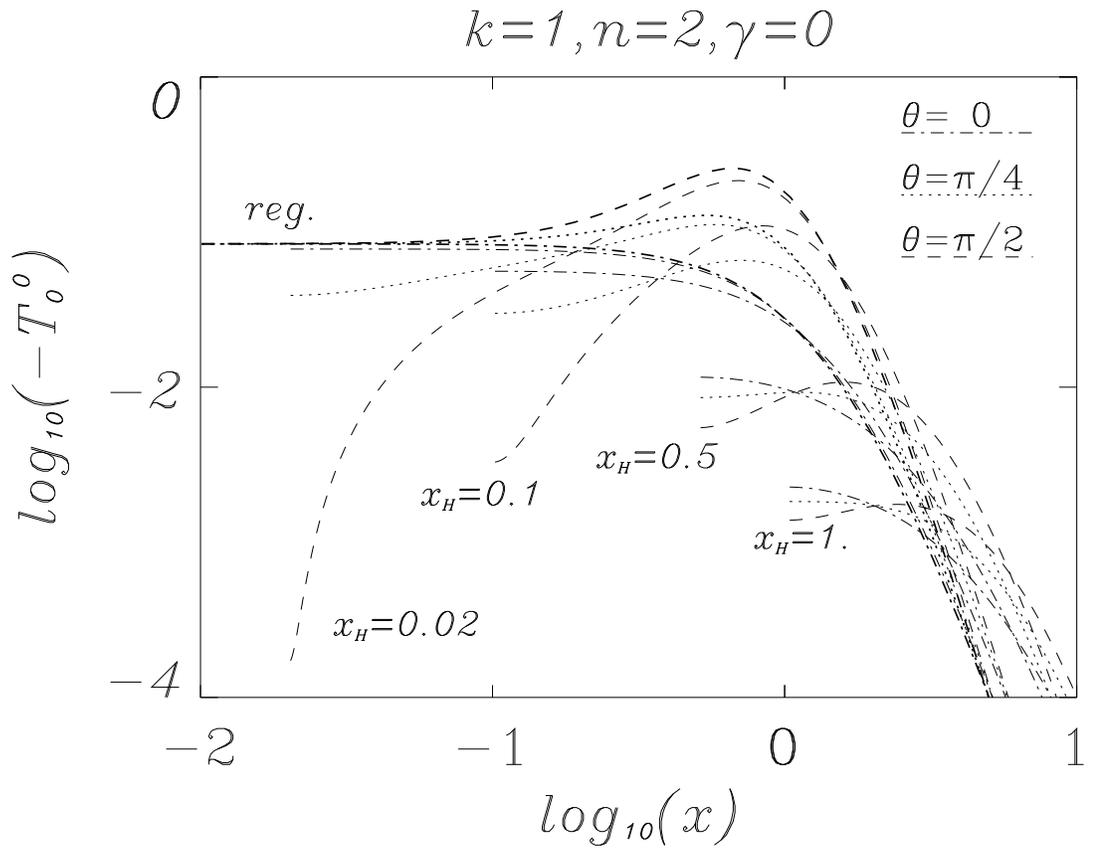}}
\caption{\label{bh1}
The energy density $\epsilon=-T_0^0$ is shown as a function 
of the dimensionless coordinate $x$ for the angles 
$\theta=0$ (dash-dotted),
$\theta=\pi/4$ (dotted)
and $\theta=\pi/2$ (dashed)
for the EYM black hole solutions with
winding number $n=2$, node number $k=1$ and 
horizon radius $x_{\rm H}=0.02$,
$x_{\rm H}=0.1$, $x_{\rm H}=0.5$,
and $x_{\rm H}=1$,
as well as for the corresponding regular solution.
}
\end{figure}

\begin{figure}
\centering
\epsfysize=11cm
\mbox{\epsffile{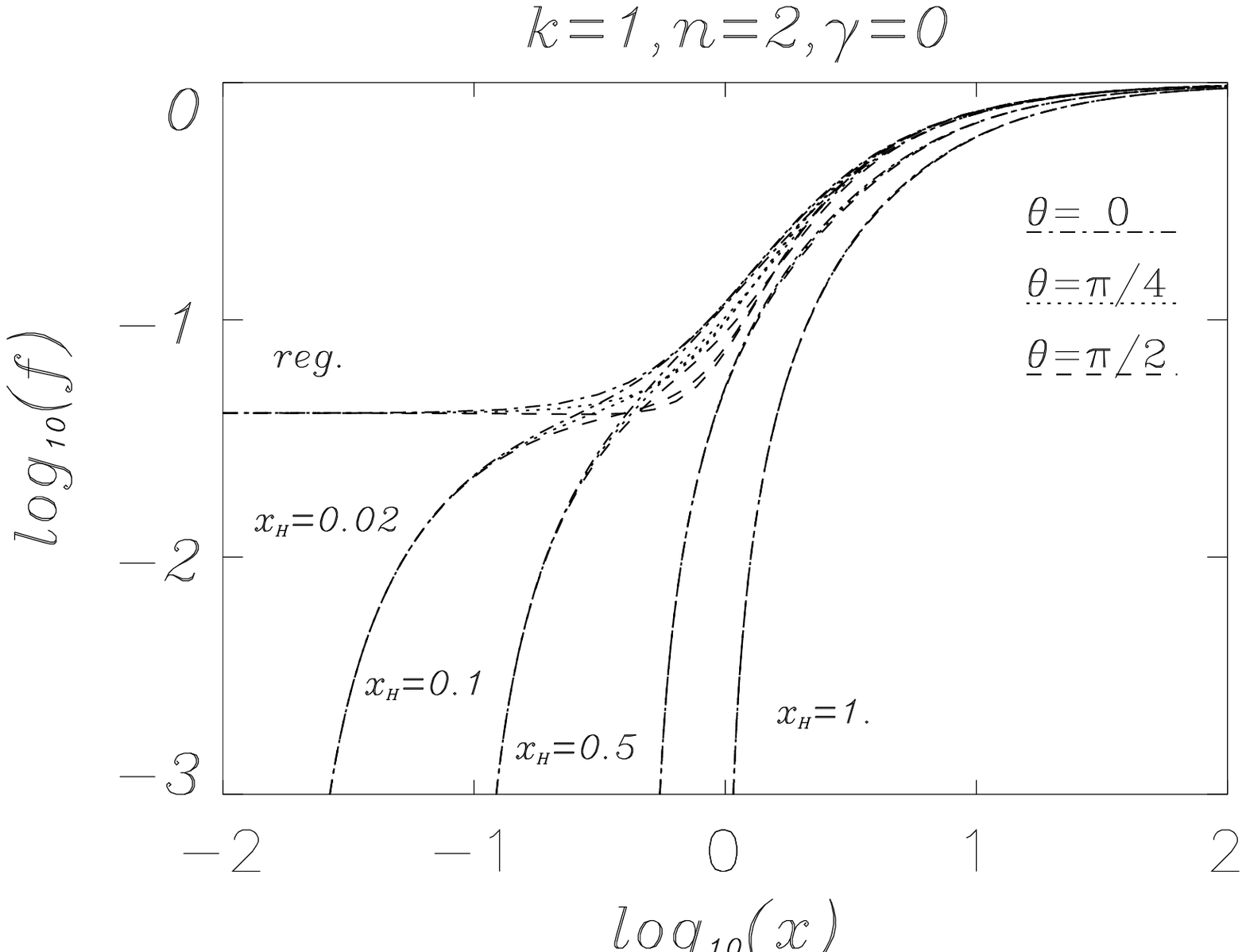}}
\caption{\label{bh2}
Same as Fig.~\ref{bh1} for the metric function $f$.
}
\end{figure}

\begin{figure}
\centering
\epsfysize=11cm
\mbox{\epsffile{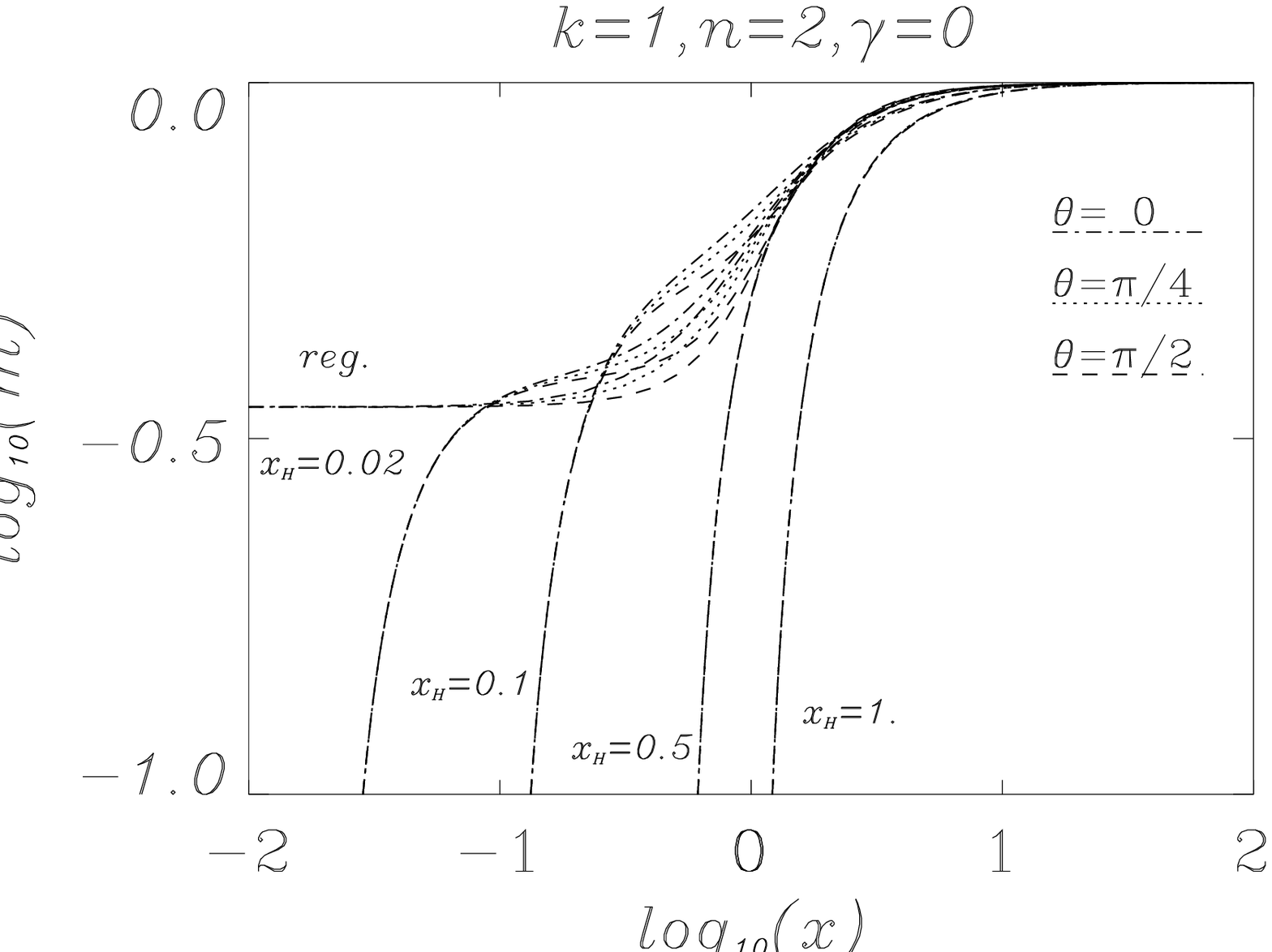}}
\caption{\label{bh3}
Same as Fig.~\ref{bh1} for the metric function $m$.
}
\end{figure}

In Fig.~\ref{bh1}
we show the energy density of the black hole solutions
with $n=2$, $k=1$, $\gamma=0$ and 
$x_{\rm H}=0.02,\ 0.1, \ 0.5$ and 1 for several angles.
The figure thus illustrates how, with decreasing horizon,
the energy density tends towards the energy density 
of the corresponding regular solution, which is also shown.
For small values of $x_{\rm H}$, the energy density of the black
hole solutions has a strong peak on the $\rho$-axis
(away from the horizon), just like
the energy density of the corresponding regular solutions.
As the value of $x_{\rm H}$ increases
the energy density changes its shape.
For larger values of $x_{\rm H}$
the maximum occurs on the $z$-axis at the horizon.
With increasing $x_{\rm H}$
the energy density of the matter fields becomes less important.
Therefore the metric functions become more spherical.
This is demonstrated in Figs.~\ref{bh2} and \ref{bh3}
for the metric functions
$f$ and $m$, respectively, for the same parameters.
($l$ looks similar to $m$.)
For fixed $n$ and $x_{\rm H}$ and increasing $k$, 
the angular dependence of the solutions decreases,
whereas for fixed $k$ and $x_{\rm H}$ and increasing $n$, 
the angular dependence of the solutions increases.

The situation is analogous for finite dilaton coupling constant
$\gamma$. In particular,
the angle-dependence of the dilaton function also decreases
with increasing $x_{\rm H}$ for fixed $n$ and $k$.
For small values of $x_{\rm H}$
the dilaton function $\varphi$ of the black hole solutions
looks similar to the dilaton function of the corresponding regular 
solutions \cite{kk2}.
However, it has a small angle-dependence at the horizon.

\begin{figure}
\centering
\epsfysize=11cm
\mbox{\epsffile{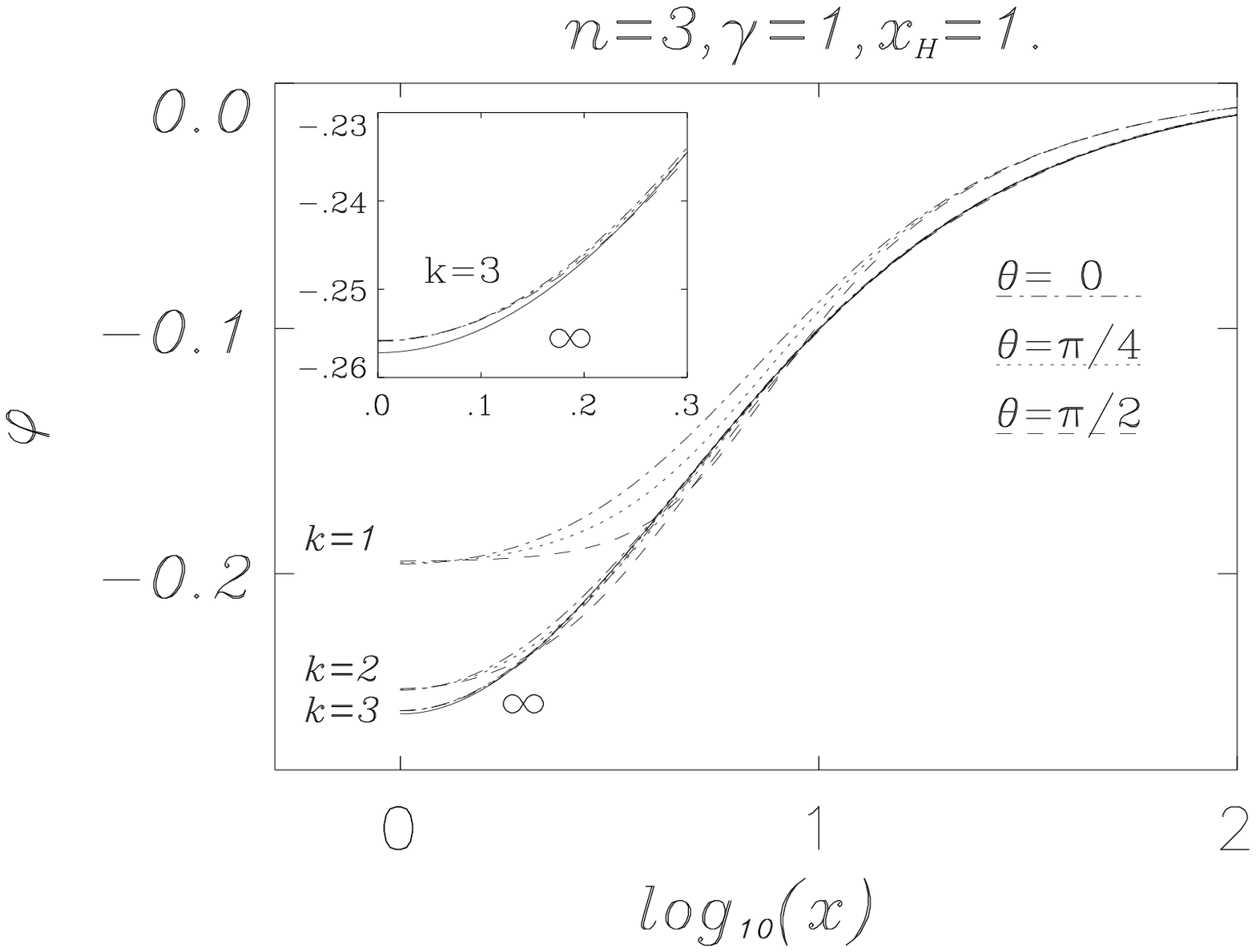}}
\caption{\label{bh4}
The dilaton function $\varphi_k(x,\theta)$ 
is shown as a function
of the dimensionless coordinate $x$ for the angles 
$\theta=0$ (dash-dotted),
$\theta=\pi/4$ (dotted)
and $\theta=\pi/2$ (dashed)
for the EYMD solutions with
winding number $n=3$, node numbers $k=1$, $k=2$ and $k=3$,
horizon radius $x_{\rm H}=1$
and dilaton coupling constant $\gamma=1$.
Also shown is the limiting function
$\varphi_\infty(x)$ (solid).
}
\end{figure}

For fixed $n$ and $\gamma$ and increasing $k$,
the regular static axially symmetric solutions form sequences,
which tend to ``extremal'' EMD solutions \cite{emd}
with $n$ units of magnetic charge and the same value of $\gamma$
\cite{kk2}.
Here we observe a similar convergence for the black hole
solutions 
for fixed $n$, $\gamma$ and $x_{\rm H}$ and increasing $k$
\cite{kks3,kknew}.
For finite values of $\gamma$,
the sequences of static axially symmetric black hole solutions
converge to EMD black hole solutions
with $n$ units of magnetic charge, the same value of $\gamma$
and the same $x_{\rm H}$.
For $\gamma=0$ the limiting solutions are 
the corresponding Reissner-Nordstr\o m solutions.
This convergence is demonstrated in Fig.~\ref{bh4} for the dilaton
function $\varphi$ for $n=3$, $\gamma=1$ and $x_{\rm H}=1$
as well as in Table~1 for the mass, dilaton charge and temperature.
More details of the solutions will be given elsewhere \cite{kknew}.

\section{Discussion}

The black hole solutions constructed here for EYM and EYMD theory
are of a novel type.
They are asymptotically flat, static 
and possess a regular event horizon \cite{ks}.
However, they are only axially symmetric
with angle-dependent fields on the horizon.
Whereas the energy density of the regular solutions
has a torus-like shape, due to a strong peak on the $\rho$-axis
away from the origin \cite{kk2},
the energy density of the black hole solutions has
a similar shape only for small values of $x_{\rm H}$.
For larger values of $x_{\rm H}$
the shape of the energy density changes, and
a strong peak develops on the $z$-axis at the horizon.
The shape of the energy density
becomes ellipsoidal and the angle-dependence diminishes.

There is all reason to believe, that these static
axially symmetric EYMD and EYM black hole solutions are unstable.
But we expect analogous solutions in EYMH theory \cite{ewein}
and ES theory,
corresponding to black holes inside axially symmetric multimonopoles
and multiskyrmions, respectively,
and for $n=2$ these axially symmetric solutions
should be stable \cite{ewein,bc}.
In contrast, the stable black hole solutions
with higher magnetic charges (EYMH) or higher baryon numbers (ES)
should not correspond to such axially symmetric solutions with $n>2$.
Instead these stable solutions should have much more complex shapes,
exhibiting discrete crystal-like symmetries \cite{ewein,bc}.
Analogous but unstable black hole solutions of this kind
should also exist in EYM and EYMD theory.

We would like to thank the RRZN in Hannover for computing time.

\newpage
\begin{table}[p!]
\begin{center}
\begin{tabular}{|c|cc|cc|} \hline
\multicolumn{1}{|c|} { $ $ }&
\multicolumn{2}{ c|} {  EYM$(\gamma=0)$ }&
\multicolumn{2}{ c|}  { EYMD$(\gamma=1)$ } \\
 \hline
 $k/n$      &  $2$  &   $3$  &  $2$  & $3$  \\
 \hline
\multicolumn{1}{|c|} { $ $ }&
\multicolumn{4}{ c|}  {$\mu$ } \\
 \hline 
 $1$        &  $2.652  $ &   $3.103  $ &  $2.565  $ & $2.897  $ \\ 
 $2$        &  $2.807  $ &   $3.489  $ &  $2.710  $ & $3.229  $ \\  
 $3$        &  $2.826  $ &   $3.586  $ &  $2.730  $ & $3.324  $ \\  
 $\infty$   &  $2.828  $ &   $3.606  $ &  $2.732  $ & $3.345  $ \\
 \hline
\multicolumn{1}{|c|} { $ $ }&
\multicolumn{4}{ c|}  {$T$ } \\
 \hline 
 $1$        &  $0.0149  $ &   $0.0126  $ &  $0.0158  $ & $0.0144  $ \\ 
 $2$        &  $0.0138  $ &   $0.0107  $ &  $0.0148  $ & $0.0126  $ \\  
 $3$        &  $0.0137  $ &   $0.0102  $ &  $0.0146  $ & $0.0120  $ \\  
 $\infty$   &  $0.0137  $ &   $0.0101  $ &  $0.0146  $ & $0.0119  $ \\
\hline
\multicolumn{1}{|c|} { $ $ }&
\multicolumn{4}{ c|}  {$D$ } \\
 \hline 
 $1$        &  $0.  $ &   $0.  $ &  $0.623  $ & $0.974  $ \\ 
 $2$        &  $0.  $ &   $0.  $ &  $0.717  $ & $1.264  $ \\  
 $3$        &  $0.  $ &   $0.  $ &  $0.731  $ & $1.332  $ \\  
 $\infty$   &  $0.  $ &   $0.  $ &  $0.732  $ & $1.345  $ \\
 \hline 
\end{tabular}
\end{center} 
\vspace{1.cm} 
{\bf Table 1}\\
The dimensionless mass $\mu$,
the temperature $T$ and the dilaton charge $D$,
of the black hole solutions
of EYM and EYMD theory for $\gamma=0$ and $\gamma=1$,
respectively, with horizon at  $x_{\rm H}=1$,
with winding numbers $n=2$ and $3$ and up to 3 nodes.
For each quantity
the corresponding limiting values are shown in the last row
(denoted by $\infty$).

\end{table}

\clearpage
\newpage

\clearpage
\newpage

\clearpage
\newpage

\clearpage
\newpage

\begin{thebibliography}{000}
\bibitem{bek}
 see e.g.~A.~E. Mayo and J.~D. Bekenstein,
 No hair for spherical black holes: charged and nonminimally coupled
 scalar field with self--interaction,
 Phys. Rev. D54 (1996) 5059.
\bibitem{hair-em}
 W. Israel,
 Event horizons in static electrovac space-times,
 Commun. Math. Phys. 8 (1968) 245;
 D.~C. Robinson,
 Uniqueness of the Kerr black hole,
 Phys. Rev. Lett. 34 (1975) 905;
 P. Mazur,
 Proof of uniqueness of the Kerr-Newman black hole solution,
 J. Phys. A 15 (1982) 3173.
\bibitem{ewein}
 S.~A. Ridgway and E.~J. Weinberg,
 Static black hole solutions without rotational symmetry,
 Phys.Rev. D52 (1995) 3440.
\bibitem{slomo}
 M.~S. Volkov and N. Straumann, 
 Slowly rotating nonabelian black holes,
 hep-th/9704026.
\bibitem{eym}
 M.~S. Volkov and D.~V. Galt'sov,
 Black holes in Einstein-Yang-Mills theory,
 Sov. J. Nucl. Phys. 51 (1990) 747;\\
 P. Bizon,
 Colored black holes,
 Phys. Rev. Lett. 64 (1990) 2844;\\
 H.~P. K\"unzle and A.~K.~M. Masoud-ul-Alam,
 Spherically symmetric static SU(2) Einstein-Yang-Mills fields,
 J. Math. Phys. 31 (1990) 928.
\bibitem{bm}
 R. Bartnik and J. McKinnon,
 Particlelike solutions of the Einstein-Yang-Mills equations,
 Phys. Rev. Lett. 61 (1988) 141.
\bibitem{strau}
 N. Straumann and Z.~H. Zhou,
 Instability of the Bartnik-McKinnon solutions
 of the Einstein-Yang-Mills equations,
 Phys. Lett. B237 (1990) 353;\\
 N. Straumann and Z.~H. Zhou,
 Instability of colored black hole solutions,
 Phys. Lett. B243 (1990) 33.
\bibitem{eymd}
 E.~E. Donets and D.~V. Gal'tsov,
 Stringy sphalerons and non-abelian black holes,
 Phys. Lett. B302 (1993) 411;\\
 G. Lavrelashvili and D. Maison,
 Regular and black hole solutions of Einstein-Yang-Mills
 dilaton theory,
 Nucl. Phys. B410 (1993) 407.
\bibitem{kk1}
 B. Kleihaus and J. Kunz,
 Axially symmetric multisphalerons in Yang-Mills-dilaton theory,
 Phys. Lett. B392 (1997) 135.
\bibitem{kk2}
 B. Kleihaus and J. Kunz,
 Static axially symmetric solutions of 
 Einstein-Yang-Mills-dilaton theory,
 Phys. Rev. Lett. 78 (1997) 2527
\bibitem{rr} 
 C. Rebbi and P. Rossi, 
 Multimonopole solutions in the Prasad-Sommerfield limit,
 Phys. Rev. D22 (1980) 2010.
\bibitem{foot1} 
 The functions $H_i$ are related to
 the functions $F_i(r,\theta)$ of refs.~\cite{kk,kk1,kk2} via
 $H_1=\sin \theta \cos \theta (F_2-F_1)$,
 $H_3=\sin \theta \cos \theta (F_4-F_3)$,
 $H_2=\cos^2\theta F_1 + \sin^2\theta F_2$,
 $H_4=\cos^2\theta F_3 + \sin^2\theta F_4$.
\bibitem{kkb}
 B. Kleihaus, J. Kunz and Y. Brihaye,
 The electroweak sphaleron at physical mixing angle,
 Phys. Lett. 273B (1991) 100;\\
 J. Kunz, B. Kleihaus, and Y. Brihaye,
 Sphalerons at finite mixing angle,
 Phys. Rev. D46 (1992) 3587.
\bibitem{kk}
 B. Kleihaus and J. Kunz, 
 Multisphalerons in the weak interactions,
 Phys. Lett. B329 (1994) 61;\\
 B. Kleihaus and J. Kunz, 
 Multisphalerons in the Weinberg-Salam theory,
 Phys. Rev. D50 (1994) 5343.
\bibitem{kknew}
 B. Kleihaus and J. Kunz, in preparation.
\bibitem{rh1}
 The perturbative calculations in EYMH theory \cite{ewein}
 yield (to leading order)
 a spherical horizon even for perturbative black hole solutions 
 with no rotational symmetry.
\bibitem{rh2}
 The horizon of the axially symmetric Kerr solution 
 in Boyer-Lindquist coordinates
 also resides at a surface of constant $r$.
\bibitem{wein}
 S. Weinberg,
 Gravitation and Cosmology
 (Wiley, New York, 1972)
\bibitem{schoen}
 W. Sch\"onauer and R. Wei\ss , 
 J. Comput. Appl. Math. 27, 279 (1989) 279;
 M. Schauder, R. Wei\ss\ and W. Sch\"onauer, 
 The CADSOL Program Package,
 Universit\"at Karlsruhe, Interner Bericht Nr. 46/92 (1992).
\bibitem{emd}
 G.~W. Gibbons and K. Maeda,
 Black holes and membranes in higher-dimensional
 theories with dilaton fields,
 Nucl. Phys. B298 (1988) 741;\\
 D. Garfinkle, G.~T. Horowitz and A. Strominger,
 Charged black holes in string theory,
 Phys. Rev. D43 (1991) 3140.
\bibitem{kks3}
 B. Kleihaus, J. Kunz and A. Sood,
 SU(3) Einstein-Yang-Mills-dilaton sphalerons and black holes,
 Phys. Lett. B374 (1996) 289;
 B. Kleihaus, J. Kunz and A. Sood,
 Sequences of Einstein-Yang-Mills-dilaton black holes,
 Phys. Rev. D54 (1996) 5070.
\bibitem{ks}
 The Kretschmann scalar $R^{\mu\nu\alpha\beta}R_{\mu\nu\alpha\beta}$
 is finite at the horizon.
\bibitem{bc}
 E. Braaten, S. Townsend and L. Carson,
 Novel structure of static multisoliton solutions
 in the Skyrme model,
 Phys. Lett. B235 (1990) 147.
 
\end{thebibliography}
\end{document}